\documentclass{article}
\usepackage{spconf,amsmath}
\usepackage{hyperref}       
\usepackage{url}            
\usepackage{booktabs}       
\usepackage{amsfonts}       
\usepackage{microtype}      
\usepackage{lipsum}		    
\usepackage{graphicx}
\usepackage{amsthm,amssymb,latexsym}
\usepackage{color}
\usepackage{physics}
\usepackage{setspace}

\title{Quantum Federated Learning with Quantum Networks}

\name{Tyler Wang$^1$, Huan-Hsin Tseng$^2$, Shinjae Yoo$^2$}
\address{$^1$ Department of Computer Science, Stony Brook University, Stony Brook, NY 11794\\
$^2$ Computational Science Initiative, Brookhaven National Laboratory, Upton, NY 11793}

\begin{document}
\maketitle

\begin{abstract}
    A major concern of deep learning models is the large amount of data that is required to build and train them, much of which is reliant on sensitive and personally identifiable information that is vulnerable to access by third parties. Ideas of using the quantum internet to address this issue have been previously proposed, which would enable fast and completely secure online communications. Previous work has yielded a hybrid quantum-classical transfer learning scheme for classical data and communication with a hub-spoke topology. While quantum communication is secure from eavesdrop attacks and no measurements from quantum to classical translation, due to no cloning theorem, hub-spoke topology is not ideal for quantum communication without quantum memory. Here we seek to improve this model by implementing a decentralized ring topology for the federated learning scheme, where each client is given a portion of the entire dataset and only performs training on that set. We also demonstrate the first successful use of quantum weights for quantum federated learning, which allows us to perform our training entirely in quantum.
\end{abstract}

\section{Introduction}
    Federated learning (FL) has recently emerged as a major tenet of machine learning (ML), particularly deep learning (DL) due to the growing privacy concerns associated with the dependency of these models on data. Although there are a handful of public open-source datasets available to researchers, the majority of successful ML/DL models utilize data collected from users' devices (e.g., Google searches, financial records, etc.). Such information should not be easily accessible to malicious third parties, who, if given access to the central computing infrastructure, can thereby compromise and leak high-stakes data. Federated learning helps to mitigate this concern by focusing on decentralizing computer architecture so that users are able to upload their individual data to the cloud without the risk of third-party access. In other words, each client's data is independent of and private to outsiders; individuals are able to train their own models and contribute to the global model without revealing the content of the model itself. 

    The concept of FL offers great promise when integrated with quantum machine learning (QML) to create QFL. One of the key benefits to QML/QFL is the no-cloning theorem, which does not allow for the existence of two identical but unknown quantum states. This adds to the security of client data: if the network were penetrated by unauthorized third party, they would not be able to copy down the information, as its state would change immediately after. 
    
    Prior research on the area of quantum federated learning (QFL) has shown its feasibility and advantages in training large machine learning models. The work in ~\cite{chen2021federated} demonstrated a hybrid quantum-classical learning scheme, extracting data from a pre-trained classical neural network and encoding it into a quantum state. These data were then used in the QFL model to make predictions for image classification. Additionally, the authors of ~\cite{chehimi2022quantum} develop a QFL framework that utilizes quantum convolutional neural networks for classification tasks. They were also able to successfully generate the first purely quantum federated dataset, which they train using QFL. 

\begin{figure}[h]
        \centering
        \includegraphics[width=\columnwidth, height=1.5in]{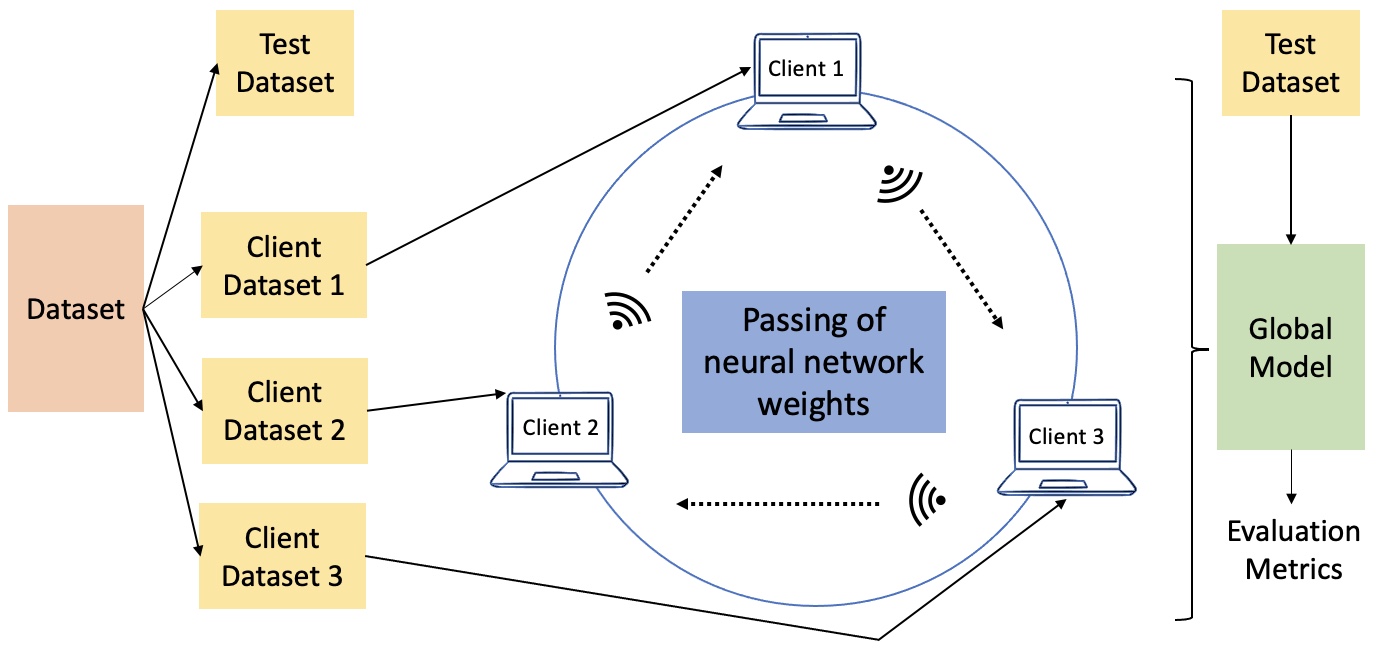}
        \caption{Decentralized QFL framework using a ring network to organize the clients. The training data is divided among each client, who each trains their portion of the data separately, updating and passing weights to the next client until all training rounds have finished. At the end of training, we use the last client's model (in this example, client 3) and its parameters to calculate the metrics of our model against the test dataset - in this case, the accuracy.}
        \label{fig:general-qfl-structure}
    \end{figure}

    We build on this work by providing the first implementation of quantum weights in QFL, which are communicated using quantum teleportation methods between clients. We evaluate the performance of this model against the quantum federated learning and classical federated learning (CFL) models which use classical weights and classical communication methods. A general outline of our decentralized QFL network is provided in Figure ~\ref{fig:general-qfl-structure}.

    The rest of this paper is organized as follows: Section~\ref{Sec: setup} describes the experiment setup and the federated learning model in detail. Section ~\ref{Sec: communication} discusses the schemes used to implement quantum weights into training. Section~\ref{Sec: results} shows the performance of all three FL models. Section~\ref{Sec: discussion} contains further discussions about these results. We draw conclusions in Section~\ref{Sec: conclusion}.

\section{Problem Setup}\label{Sec: setup}
    
    \subsection{Federated Learning Topology}
\begin{figure}[h]
        \begin{minipage}[b]{0.46\linewidth}
          \centering
          \centerline{\includegraphics[width=\linewidth]{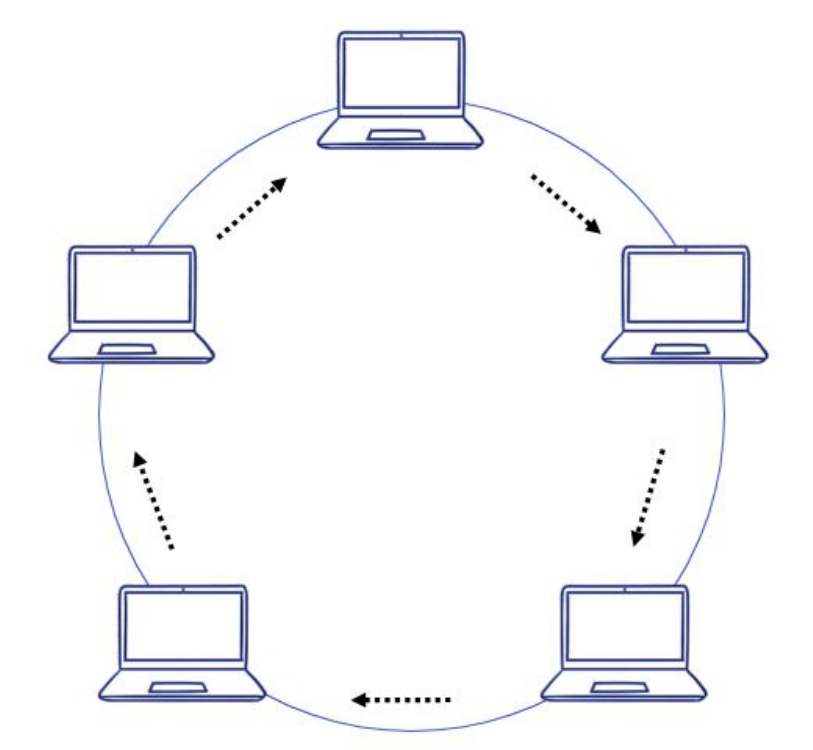}}
          \centerline{(a) Ring Topology}\medskip
        \end{minipage}
        \hspace{.7cm}
        \begin{minipage}[b]{0.46\linewidth}
          \centering
          \centerline{\includegraphics[width=\linewidth]{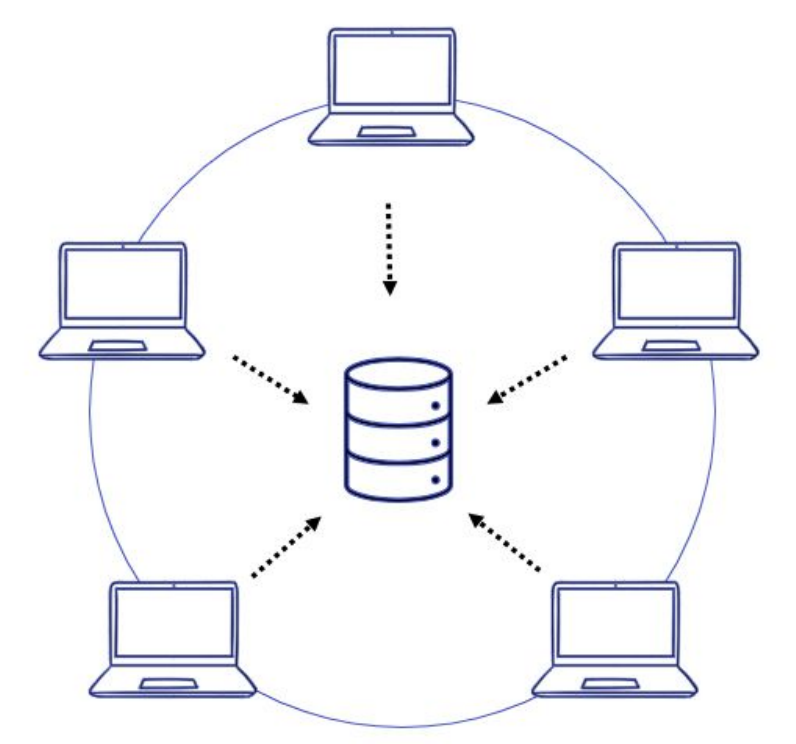}}
          \centerline{(b) Hub-Spoke Topology}\medskip
        \end{minipage}
        \caption{Topology of a ring network (a) compared to a hub-spoke network (b), both with five client nodes. Note that the hub-spoke topology contains an additional central hub node to which clients pass their parameters after training, while in the ring topology, clients simply exchange parameters with other clients.}
        \label{fig:fl-architectures}
        \end{figure}

        For both the classical and quantum neural networks that we implement, we use what is known as a \textit{ring network topology}, which has been shown to yield better accuracy of federated learning in neural networks (\cite{wang2022decentrfl}\cite{han2023ringffl}\cite{li2022fedhisyn}\cite{xu2023ring}\cite{lee2021tornadoaggregate}). In this ring, each client only obtains a subset of the entire dataset and performs their training only on that set. The updated parameters of the training are passed to the next client, and so on. Once the training has finished, the final client's model is used to compute the accuracy against the test dataset. This is different from hub-spoke network topology, where the global model is shared with several clients. After the training round is complete, each clients' model is uploaded to the central server, which performs an averaging process to update the global parameters. The global model is then distributed among the clients, and is later used in accuracy calculations. These two network architectures are contrasted in Figure ~\ref{fig:fl-architectures}.

\subsection{Quantum Neural Networks}

Quantum neural networks (QNN), or variational quantum circuits (VQC), contains circuit parameters that can be updated using gradient-based methods. Recent research underscores the superior expressiveness of VQCs compared to classical neural networks \cite{sim2019expressibility, lanting2014entanglement, du2018expressive, abbas2021power}. Moreover, VQCs exhibit efficacy in training with smaller datasets, as demonstrated by Caro et al. \cite{caro2022generalization}. VQC applications span various domains in QML, including classification, reinforcement learning, natural language processing, and sequence modeling \cite{mitarai2018quantum, qi2023theoretical, chen2021end, chehimi2022quantum, chen2020variational, chen2023quantum, yang2020decentralizing, yang2022bert, di2022dawn, li2023pqlm, chen2020quantum}.

The VQC contains three essential parts: a quantum encoder that provides a routine for encoding classical data into a quantum state as well as a variational quantum circuit block with the learnable parameters. The final part of the VQC consists of measuring the qubits, which gives classical values that can further be processed in other quantum or classical components during the training and inference process. 
%

\subsubsection{Quantum Encoder}
To train our QFL scheme on a classical dataset, it is essential to find a way to transform the classical data vectors into quantum states. We implement the \textit{variational encoding} method, where the input values are used as rotation angles in quantum rotational gates. For this work, we use $R_{x}$ gates.

    \begin{figure}[h]
        \centering
        \includegraphics[width=\columnwidth]{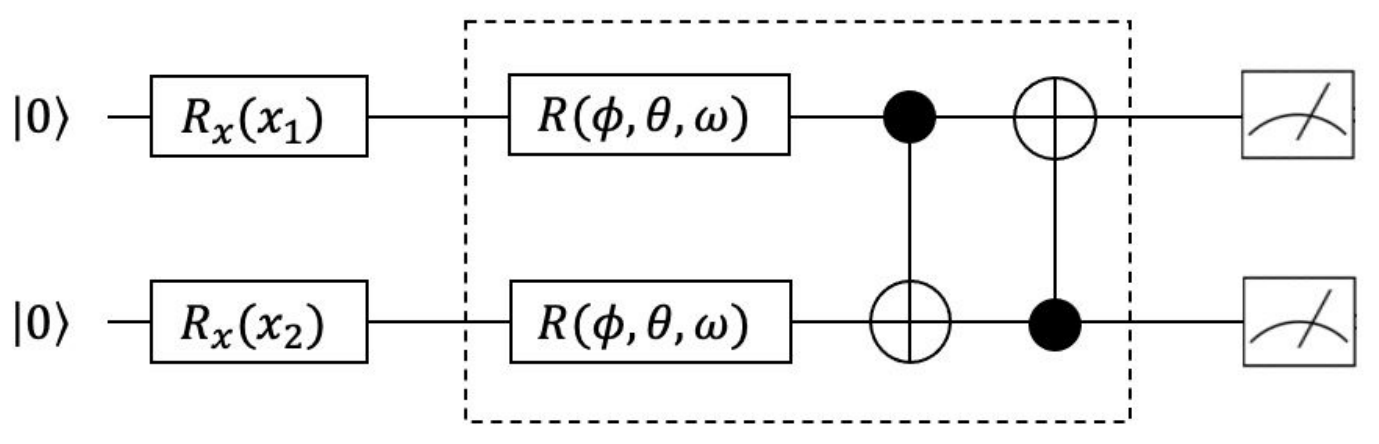}
        \caption{Architecture for the variational quantum circuit. The encoding layer of the VQC consists of single-qubit rotational $R_{x}$ gates, whose rotation angles are given by the input data. The grouped box represents the variational quantum layer, which consists of general parametrized unitary gates $R(\phi, \theta, \omega)$. These parameters $\phi, \theta,$ and $ \omega$ are subject to the optimization procedures. Finally, the quantum measurement component outputs the Pauli-Z expectation values of our qubits.}
        \label{fig:vqc}
    \end{figure}
    
\subsubsection{Quantum Layer}
    The variational layer is the learnable block of the VQC which contains the learnable parameters, or weights, to mimic the classical neural network. This part consists of CNOT gates between each pair of neighboring qubits to entangle the quantum states followed by a series of single qubit unitary gates parametrized by $\phi$, $\theta$, and $\omega$, see Figure ~\ref{fig:vqc}. After these operations have been applied, measurements are taken on the two qubits.

\subsubsection{Gradient Calculation Using Parameter Shift}\label{Subsubsec: gradient}
Given a Hermitian operator $B$ (observable), a fixed encoder $E(x)$ and a unitary gate $R_i \in U(2)$ of free parameters $\theta_i \in \mathbb{R}^3$, we can construct a scalar-valued quantum circuit function defined by
    \begin{equation}\label{E: 1-gate func}
    \begin{aligned}
        f(x, \theta_{i}) &= \bra{0} E^\dagger(x) \, R_{i}^\dagger(\theta_{i}) \, B  \, R_{i}(\theta_{i}) \, E(x) \, \ket{0} \\ 
        &:= \bra{x}R_{i}^\dagger(\theta_{i})\, B \, R_{i}(\theta_{i})\ket{x}
        \end{aligned}
    \end{equation}
where $x$ is originally a classical data point of vector forms converted into a quantum state denoted as $\ket{x} :=E(x) \ket{0}$ by a fixed embedding operator $E(x)$.
Denoting $M_{\theta_{i}}(B):= R_{i}^\dagger(\theta_{i}) \, B \, R_i (\theta_{i})$, the gradient of the scalar function Eq.~(\ref{E: 1-gate func}) with respect to $\theta_i$ can be written as:
\begin{equation}
    \nabla_{\theta_{i}}f(x,\theta_{i}) = \bra{x}\nabla_{\theta_{i}}M_{\theta_{i}}(B)\ket{x}
\end{equation}

For multiple parameterized gates, consider a product of unitary operators $R_{i}(\theta_{i})$ with learnable parameters $\theta_i$:
\begin{equation}\label{E: n-qubit circuit gate}
    R(x,\theta) := R_{n}(\theta_{n}) R_{n-1}(\theta_{n-1}) ... R_{i}(\theta_{i}) ...  R_{1}(\theta_{1}) R_{0}(x)
\end{equation}
Then we may define a scalar function of multiple parameterized gates by $f(x, \theta) = \bra{0} R^{\dagger}(x,\theta) \, B \, R(x,\theta) \ket{0}$ whose gradient is given by
\begin{equation}
     \nabla_{\theta_{i}}f(x, \theta) = \bra{\phi_{i-1}}\nabla_{\theta_{i}} M_{\theta_{i}}(B_{i+1}) \ket{\phi_{i-1}}
\end{equation}
where 
\[
\begin{aligned}
&M_{\theta_i}(B_{i+1}) := R^{\dagger}_i(\theta_i) \, B_{i+1} R_i(\theta_i)\\
&B_{i+1} := R_{i+1}^{\dagger}(\theta_{i+1}) \cdots R_n^{\dagger}(\theta_n)  B \,  R_{n}(\theta_{n}) \cdots R_{i+1}(\theta_{i+1})
\end{aligned}
\]

and a relationship is to be satisfied:
\begin{equation}
\nabla_{\theta_i} M_{\theta_i}(B) = c \left( M_{\theta_i + s}(B) - M_{\theta_i - s}(B) \right)
\end{equation}
This is called the \textit{parameter-shift rule} for updating the parameters $\theta_i$.



\section{Quantum Communication}\label{Sec: communication}
\begin{figure}[htbp]
\centering
\includegraphics[width=\columnwidth]{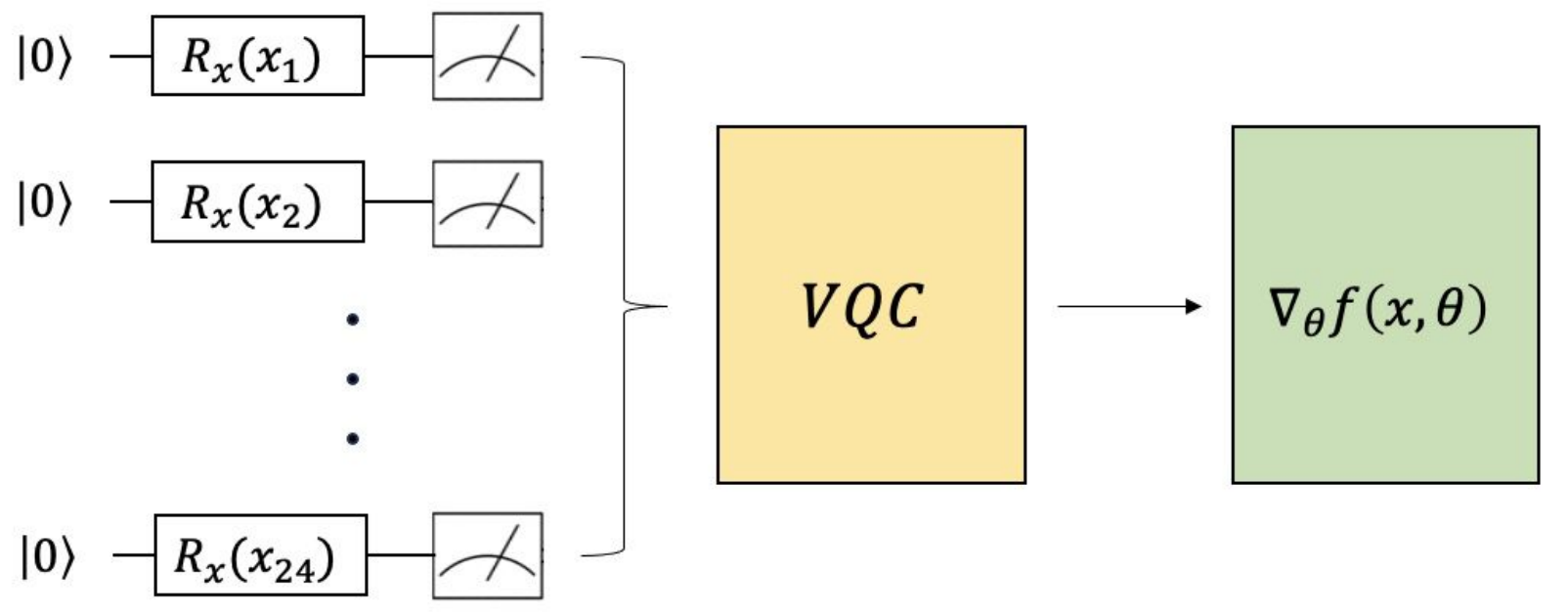}
\caption{Use of quantum weights in quantum federated learning. The weights are encoded using $R_{x}$ gates, then used in the VQC and updated with parameter-shift.}
\label{fig:quantum-weights}
\end{figure}

Within the majority of QFL frameworks, there is \textit{classical} communication of weights, both from client to client and client to server. In this work, we demonstrate the \textit{quantum} communication of quantum weights throughout the QFL network. To accomplish this, we first transform the weights into quantum states using a similar encoding process as for the input data of our neural network, obtaining the expectation values of the qubits. These values are then used as the weights in our neural network, and updated using the parameter-shift rule derived in Section~\ref{Subsubsec: gradient}. Then, we simulate quantum teleportation to transport the data to the next client device in the ring. This scheme is presented in Figure ~\ref{fig:quantum-weights}.

\section{Results}\label{Sec: results}
\begin{figure}[htbp]
\centering
\begin{minipage}[h]{1.0\linewidth}\label{fig:cfl-results}
\centering
\centerline{(a) CFL with Classical Weights}\medskip
\centerline{\includegraphics[width=\linewidth]{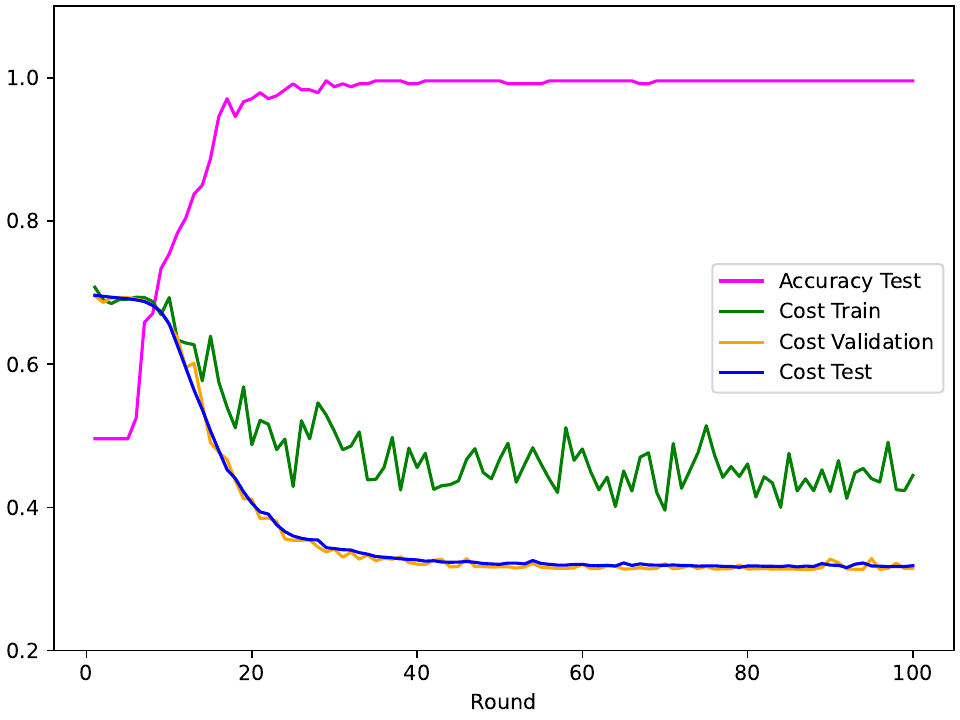}}
\end{minipage}
\hspace{.7cm}
\begin{minipage}[h]{1.0\linewidth}\label{fig:qfl-cc-results}
\centering
\centerline{(b) QFL with Classical Weights}
\centerline{\includegraphics[width=\linewidth]{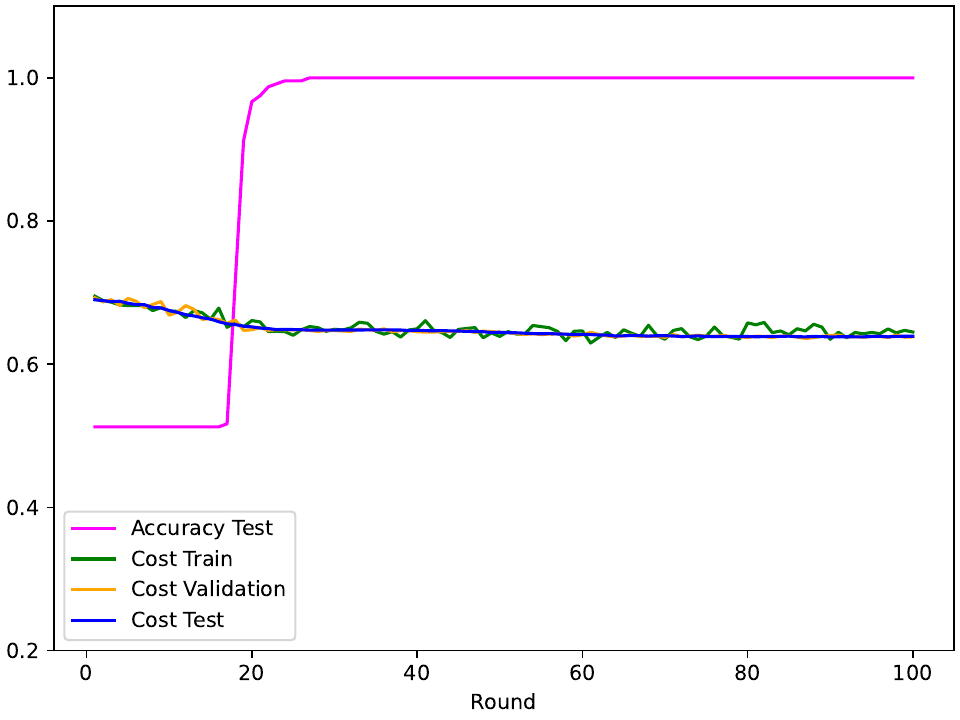}}
\end{minipage}
\begin{minipage}[h]{1.0\linewidth}\label{fig:qfl-qc-results}
\centering
\centerline{(c) QFL with Quantum Weights}
\centerline{\includegraphics[width=\linewidth]{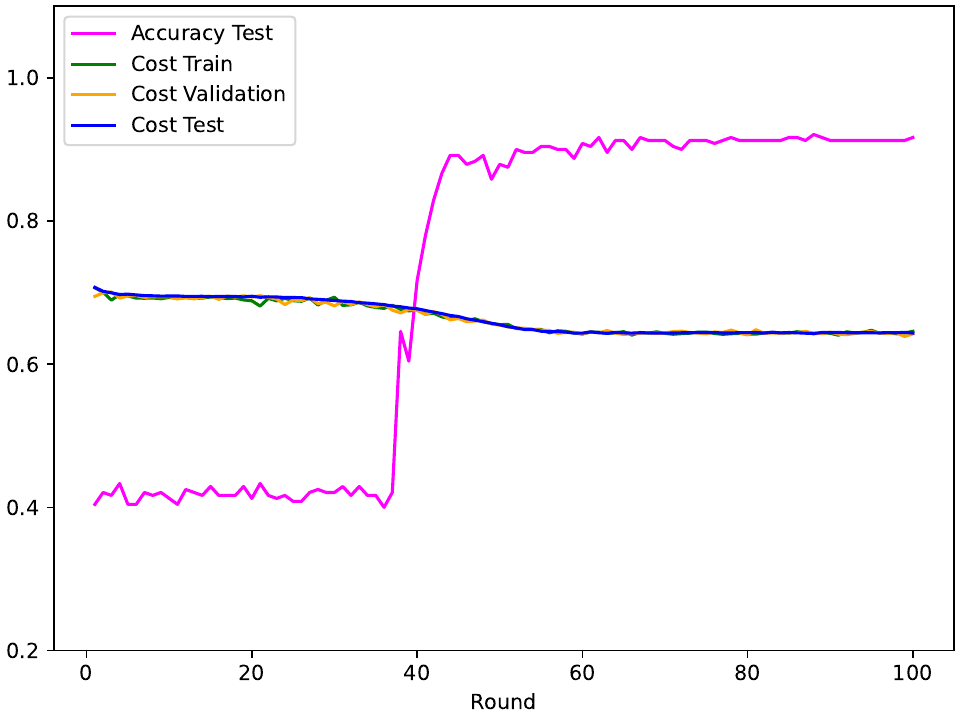}}
\end{minipage}
\caption{Results.}
\label{fig:cfl-qfl-results}
\end{figure}

We perform binary classification of the make-circles dataset, a toy dataset what is particularly useful for visualizing classification algorithms. We have in total 1200 data points, with 960 training data and 240 testing data. Once one client completes their training, they communicate their updated weights to the next client in the ring; we continue until all training has finished. Once this has been completed, we use the final client model in the training to gauge the accuracy of our model against the testing data. We run the models each for 100 rounds of training, with 5 local epochs for each client. We compare the performance of the QFL model using classical communication against the same classical architecture on the same dataset. This same procedure is repeated for the quantum communication QFL model. Our results are shown in Figure ~\ref{fig:cfl-qfl-results}.

\section{Discussion}\label{Sec: discussion}

\subsection{Decentralization}
The ring topology offers much more security of data than does hub-spoke topology. As mentioned by the authors of \cite{chen2021federated}, without the presence of a central node, we lower the risk of external access to the entire dataset. If an unauthorized third party were to penetrate a client in the ring, they would only gain access to a portion of the clients' data, rather than the whole.

This network structure also reduces the error of our quantum models. With quantum devices in the noisy intermediate-scale quantum era, error increases significantly as the number of qubits used on a single device increases. Since we use multiple quantum devices, we spread out the number of qubits used in our QFL model, thereby reducing the amount of error that is produced.

\subsection{Quantum Weights}
The most important contribution of our work is the implementation of quantum weights in QFL. This has significant impact for future developments of quantum models and quantum internet, as it removes the overlap between the quantum and classical realms. It also reduces the overhead that is produced as a result of converting between quantum and classical computational bases.

\subsection{Future Research}

\subsubsection{Quantum Networking}
    We only mimic the use of quantum teleportation in our quantum communication model; our techniques, while feasible, do not truly demonstrate teleportation of quantum weights. We plan to move our model structure to the a quantum internet simulation framework, which allows for pure quantum teleportation of quantum states between different clients. The authors of the work in ~\cite{sisodia2018comment} show a secure way to transport six-qubit states by using two Bell states as a quantum channel between clients, a method inspired by the work in ~\cite{rigolin2005teleportation}. This can be implemented with our model by teleporting the six weights of each layer separately to the next client in the training. However, our current model is sufficient to show that it is possible to use quantum weights in training of neural networks.
    
\subsubsection{Quantum Advantage}
    The dataset that we use in this work is very simple; because of this, it can be difficult to show a significnat quantum advantage. As seen in Figure ~\ref{fig:cfl-qfl-results}, the performance of the CFL model and QFL model with classical weights are similar, converging around round 25 and round 20 of training, respectively. We hope to replicate a similar scheme for image classification sets, such as the cats and dogs dataset, as well as MNIST and CIFAR10, with potentially greater indication of a quantum advantage.
    
\subsubsection{Differential Privacy}
    Another option is to incorporate QFL with the differential privacy model developed in \cite{watkins2023quantum}. This would ensure even greater confidentiality of client information, as we should be able to pass the entire VQC from one client to another without revealing the contents of the training data.
    
\section{Conclusion}\label{Sec: conclusion}


Our study establishes the quantum advantage of our QFL model over classical FL models using the make-circles dataset. We also demonstrate the success of a decentralized ring topology for training and introduce the first QFL model utilizing pure quantum weight communication through quantum teleportation methods. This work contributes to large-scale quantum internet development and enhances data privacy in ML models.
    

\clearpage
\begin{spacing}{0.6}
\footnotesize
\bibliographystyle{IEEEbib}
\bibliography{refs,bib/qml_examples,bib/vqc}
\end{spacing}
\end{document}